\documentclass[aps,pra,longbibliography,superscriptaddress,twocolumn,10pt,nofootinbib]{revtex4-1}
\usepackage[utf8]{inputenc}
\usepackage[english]{babel}
\usepackage{times}
\usepackage{color}
\usepackage{graphicx}
\usepackage{import}
\usepackage{amsmath}
\usepackage{braket}
\usepackage{amssymb}

\usepackage{hyperref}
\usepackage[all]{hypcap}

\usepackage{booktabs}

\AtBeginDocument{
\heavyrulewidth=.08em
\lightrulewidth=.05em
\cmidrulewidth=.03em
\belowrulesep=.65ex
\belowbottomsep=0pt
\aboverulesep=.4ex
\abovetopsep=0pt
\cmidrulesep=\doublerulesep{}
\cmidrulekern=.5em
\defaultaddspace=.5em
}

\usepackage{tikz}

\newcommand\encircle[1]{%
  \tikz[baseline= (X.base)]
    \node(X) [draw, shape=circle, inner sep=-0.5pt] {\tiny\strut#1};
}

\begin{document}
\title{Lattice Surgery on the Raussendorf Lattice}
\author{Daniel Herr}
\email{daniel.herr@riken.jp}
\affiliation{Theoretical Quantum Physics Laboratory, RIKEN Cluster for Pioneering Research, Saitama 351-0198, Japan}
\affiliation{Computational Physics, ETH Zurich, 8093 Zurich, Switzerland}
\author{Alexandru Paler}
\affiliation{Linz Institute of Technology, Johannes Kepler University, Linz, 4040, Austria}
\author{Simon J. Devitt}
\affiliation{Centre for Quantum Software \& Information (QSI), Faculty of Engineering \& Information Technology, University of Technology Sydney, Sydney, NSW 2007, Australia}
\affiliation{Turing Inc., Berkley CA 94705, USA}
\author{Franco Nori}
\affiliation{Theoretical Quantum Physics Laboratory, RIKEN Cluster for Pioneering Research, Saitama 351-0198, Japan}
\affiliation{Department of Physics, University of Michigan, Ann Arbor, MI 48109--1040, USA}

\begin{abstract}
 Lattice surgery is a method to perform quantum computation fault-tolerantly by using operations on boundary qubits between different patches of the planar code.  This technique allows for universal planar-code computation without eliminating the intrinsic two-dimensional nearest-neighbor properties of the surface code that eases physical hardware implementations. Lattice-surgery approaches to algorithmic compilation and optimization have been demonstrated to be more resource efficient for resource-intensive components of a fault-tolerant algorithm, and consequently may be preferable over braid-based logic. Lattice surgery can be extended to the Raussendorf lattice, providing a measurement-based approach to the surface code. In this paper we describe how lattice surgery can be performed on the Raussendorf lattice and therefore give a viable alternative to computation using braiding in measurement based implementations of topological codes.
\end{abstract}

\maketitle

\section{Introduction}
Fault-tolerant methods allow for quantum computation on systems that are prone to errors. The surface code is one of the most attractive choices for fault tolerance due to its nearest-neighbor interactions and its high error-threshold~\cite{PhysRevA.83.020302,Fowler2012}. For surface-code-based architectures, qubits can be implemented using various approaches with different methods of computation~\cite{equiv_approaches}. Among these are, for example, defects~\cite{Fowler2012} or twists~\cite{twist1} where computation is performed using braiding, or planar code patches where computation is performed using lattice surgery~\cite{Horsman2012}.

For many implementations of physical qubits, the surface code is the method of choice, but for linear-optics quantum computation~\cite{Knill2001,kieling_quantumcomputation,renormalization} or other hardware architectures that utilize probabilistic connections \cite{NTDS13} a measurement-based approach~\cite{one_way_comp} is the better choice. The Raussendorf lattice \cite{RHG06} is a measurement-based approach to the surface code and, thus, the two methods have the same benefits in terms of fault-tolerant thresholds, ability to perform a universal set of gates, and ability to largely detach the specifics of an algorithmic implementation from the underlying physical hardware.

While braiding has been the method of choice for performing fault-tolerant computation, recently lattice surgery has been investigated by several works~\cite{Yoder2017surfacecodetwist,herr_lattice_surgery,Herr2017,LS_interesting,PhysRevA.95.012321,LS_interface}. It has also been extended to different implementations than the surface code, such as the color-codes~\cite{ColorCode}.
For the Raussendorf lattice, however, only braiding has been investigated in depth. This paper closes this gap by describing how lattice surgery can be performed on the Raussendorf lattice.

First we give brief reviews on measurement-based quantum computation and error correction using both the surface code and the Raussendorf lattice. Furthermore, we show the translation between these two error correction methods. Then, the elementary operations using lattice surgery on the Raussendorf lattice are described.

\section{Brief Review}
In this first part we will briefly describe how computation can be performed using a special graph-state configuration called the Raussendorf lattice. We will then motivate the link between the Raussendorf lattice and surface-code quantum-computation. This will provide the necessary background in order to translate lattice surgery to the Raussendorf lattice.

\subsection{Graph States}
Graph states can be treated as resources which are used in measurement-based quantum computation~\cite{RBB03} in order to perform information processing. The name of graph states stems from an underlying undirected graph $G = \left( V,E \right) $ whose vertices $V \subset \mathbb{N}$ represent individual qubits. The set of edges $E \subset \left[ V \right]^2$ in this graph indicates entanglement between its vertices.

Given a graph, one can obtain a physical state by first initializing all qubits (vertices) to $\ket{+}$.
For each edge a controlled-phase gate has to be applied on its two vertices:
\begin{center}
  \includegraphics[width=0.5\columnwidth]{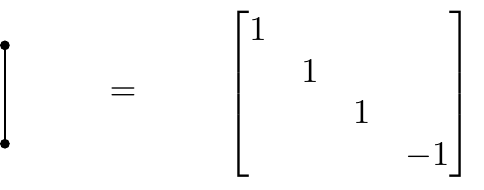}.
\end{center}

An alternative way of looking at this graph state is given in terms of stabilizer measurements. The graph state is the simultaneous eigenstate with eigenvalue $+1$ for stabilizers $S_i$ with $i \in V$ given by
\begin{equation}
S_i = X_i \prod_{j \in \text{Nbh}\left( i\right)} Z_i \qquad \qquad \text{for all} \quad i \in V.
\label{eq:stabilizer}
\end{equation}
Here, the definition of the neighborhood of a graph is given by ${\text{Nbh}\left(i\right) = \left\{ j | \left(i,j\right) \in E \right\}}$.

\subsection{Measurement-Based Quantum Computation}
  \begin{figure}
    \centering
    \includegraphics[width=\columnwidth]{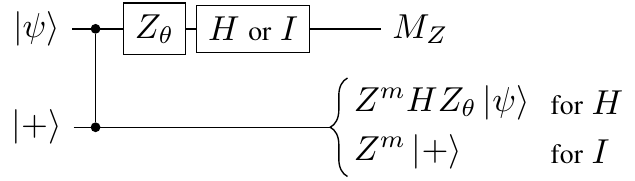}
    \caption{\label{fig:teleport1} This figure shows the circuit which teleports the initial state $\ket{\psi}$ to the next qubit in measurement-based quantum computing. The initialization step and controlled-phase gate, corresponds to the creation of a Graph state. This is followed by a measurement operation which teleports the information from one qubit to the next. Any quantum algorithm can be implemented using a sequence of these teleportation operations. Here $Z=Z_{\theta =\pi}$, $H$ is a Hadamard gate, $I$ is the identity and $M_Z$ is a measurement in $Z$-basis.}
  \end{figure}
The basic idea in measurement-based quantum computation is to to perform single-qubit measurements on this graph state. These act as teleportation operations which move the quantum state from one qubit to the next while by-product operations are applied. In Figure~\ref{fig:teleport1} the circuit for such an operation is shown. One should note that the first two steps, the initialization and the controlled-phase gate, correspond to the graph state generation. The remaining gates can be combined into a measurement in an arbitrary basis. The rotation $Z_\theta = \exp\left(i\theta Z / 2\pi\right)$ is included to perform non-clifford gates on individual qubits. Finally, measurements teleport the information from one qubit to the next. Hadamard and $Z_\theta$ operations can also be combined into a single rotated-basis measurement. Hereafter, we will always treat those operations as a single rotated-basis measurement.

However, depending on the outcome of this measurement erroneous Pauli operators are applied and need to be tracked classically. Since these measurements are probabilistic, a classical algorithm needs to track the outcomes of previous measurements and change the basis of subsequent measurements.
Furthermore, a by-product Hadamard operation is applied to the state after each teleportation operation. This causes the state to change from $Z$-basis to $X$-basis  after an odd number of measurements and changes back to $Z$-basis for an even number of measurements.

A measurement in the $Z$-basis without the application of the Hadamard operation removes the measured qubit from the lattice and no information will be teleported through this qubit. Thus, a measurement in this basis can be used to completely isolate different parts of a graph state from each other. 
Introducing defects to the lattice with $Z$-basis measurements is the main idea to implement logical qubits.

\subsection{The Raussendorf Lattice and the Surface Code}
  \begin{figure}
    \centering
    \includegraphics[width=0.9\columnwidth]{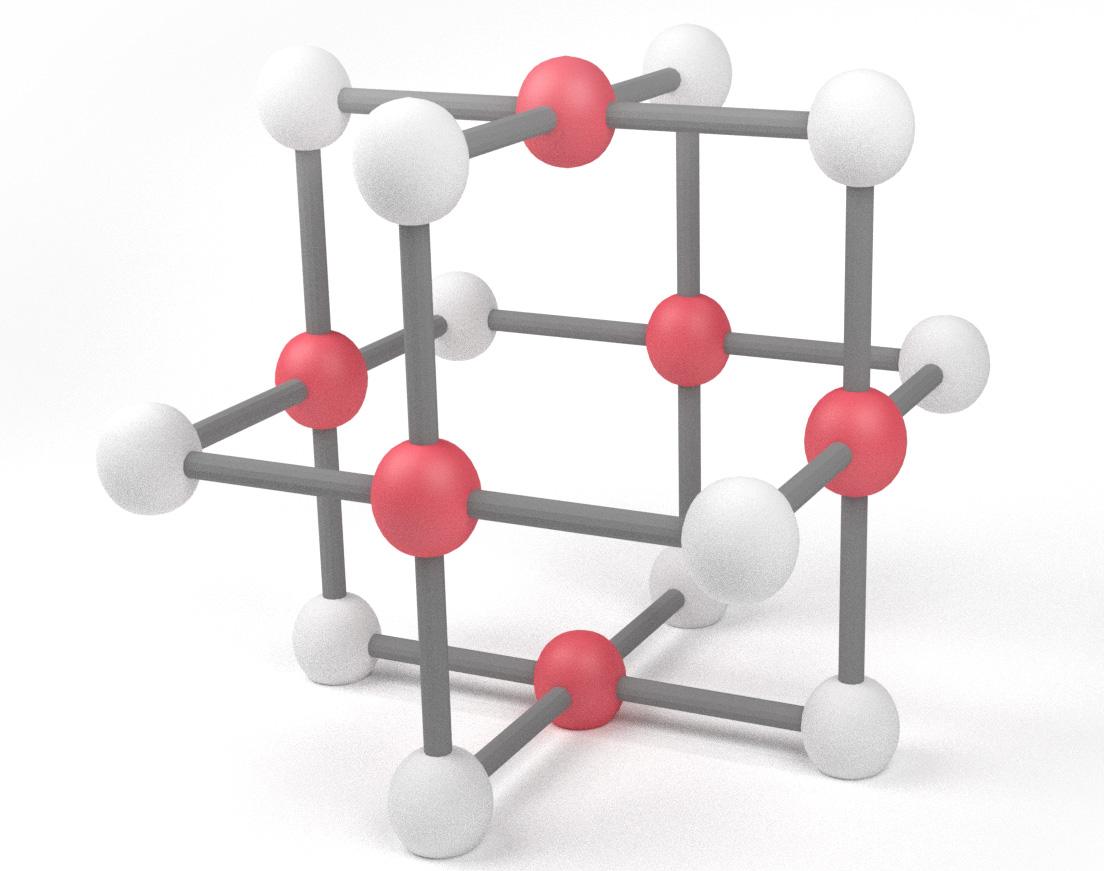}
    \caption{\label{fig:Raussendorf_unitcell} Unit cell of the Raussendorf lattice: The spheres represent individual photons and the connections between them show the entanglement given by the definition of a graph states. The vertices in the middle of the unit cell's faces are colored in red and contribute to a single parity check. The qubits colored in white are in the middle of dual-lattice faces which also correspond to parity-check operations.}
  \end{figure}

  \begin{figure}
    \centering
    \includegraphics[width=0.9\columnwidth]{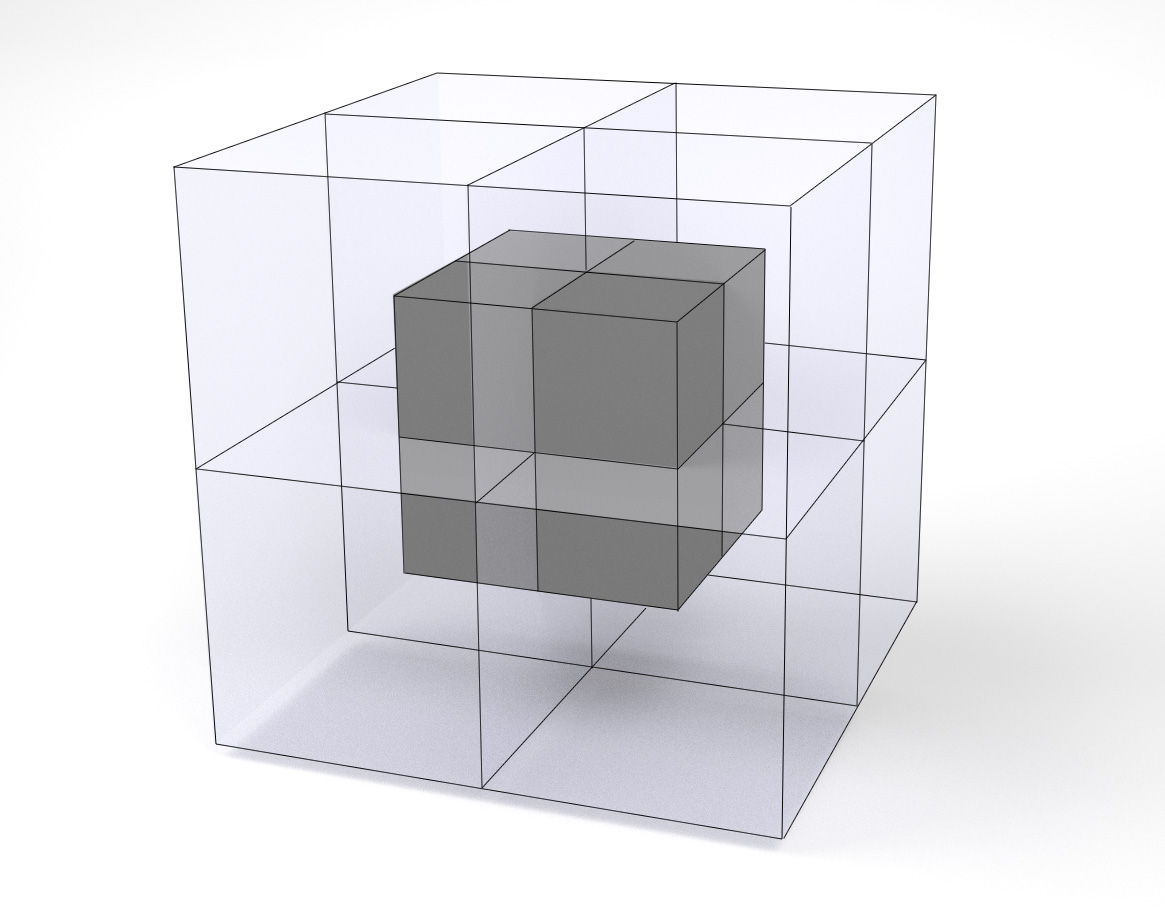}
    \caption{\label{fig:dual} The dual unit cell is represented by the dark-gray box in the center. It can be created by stacking $8$ primal unit cells (translucent boxes) together.}
  \end{figure}

The Raussendorf lattice~\cite{raussendorf_lattice} is a 3D graph-state that possesses a specific lattice structure. Figure~\ref{fig:Raussendorf_unitcell} shows a unit cell of this lattice. With the stabilizer definition given in equation~\ref{eq:stabilizer} one can see that the product of all stabilizers from the qubits colored in red corresponds to a simultaneous $X$-parity check of all the faces. Thus, we can define the $X$-stabilizer measurements for the Raussendorf lattice. This means that if no error occurred, this stabilizer measurement should give a parity of $+1$.

In order to introduce $Z$-stabilizers of the Raussendorf lattice, we need to consider the dual lattice of the Raussendorf lattice (as opposed to the primal lattice, that has been considered so far). This is a self-similar lattice that is shifted by $\left(\frac{1}{2},\frac{1}{2},\frac{1}{2}\right)$. It is visualized in Figure~\ref{fig:dual}, where the translucent boxes are given by primal unit cells and the dark box represents the dual unit cell. The faces of the dual cell correspond to the edges of the primal cells and give rise to chains of $Z$ stabilizers.

If an error occurs, two of these stabilizer measurements will show a parity of $-1$. From these syndromes one can deduce which error occurred and how to correct for it.

\subsection{Planar code}

  \begin{figure}
    \centering
    \includegraphics[width=.95\columnwidth]{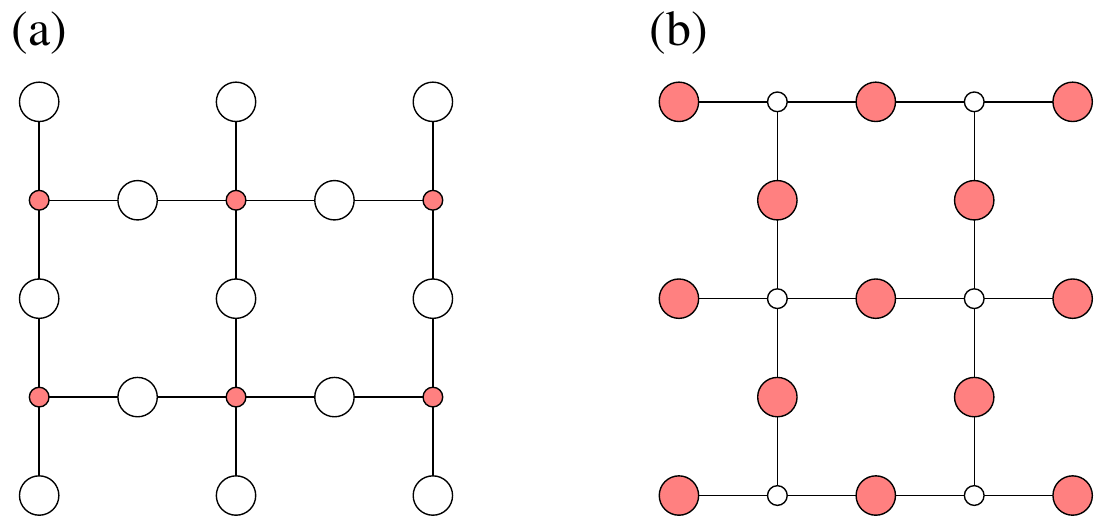}\\[0.3cm]
    \includegraphics[width=0.45\columnwidth]{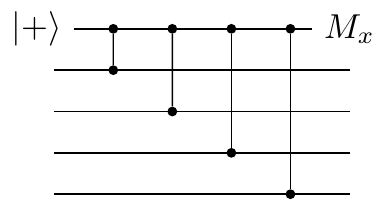}
    \hfill
    \includegraphics[width=0.47\columnwidth]{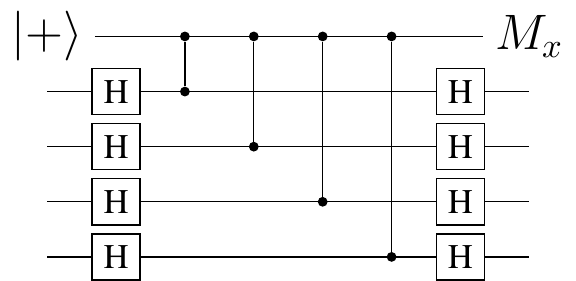}
    \caption{\label{fig:surface_code} These figures show the two stabilizer measurements of the surface code. The small circles correspond to ancilla qubits which measure the error syndromes. In (a) only $Z$ stabilizers are shown, whereas in (b) only $X$-stabilizers are shown. Merging these two stabilizers gives the complete error detection for the surface code. However, this example, where the different stabilizers are treated separately, illustrates the connection between the surface code and the Raussendorf lattice. This can be seen by looking at even (a) and odd (b) time-slices in the Raussendorf lattice. Below the lattice depictions, a single stabilizer measurement between the ancillary qubit $\ket{+}$ (top) and its surrounding data qubits is shown for both cases.}
  \end{figure}

The planar code is a 2D error-correcting code whose syndromes are continuously measured to detect errors. The only difference to the surface code is how the boundary is treated.
A layout of the planar code can be seen in Figure~\ref{fig:surface_code}. The qubits in this figure can be divided into two categories: syndrome qubits which are continuously measured and thus help detect errors, and data qubits which store the logical state of the system.
Furthermore, there are two types of syndrome measurements in different bases.

The equivalence between surface codes and the Raussendorf lattice has already been shown in~\cite{RaussendorfandSurface1,RaussendorfandSurface2} and generalized to arbitrary Calderbank-Shor-Steane stabilizer states in~\cite{foliatedcodes}. Therefore, we will only motivate why this equivalence holds and introduce quantities that we rely on by translating lattice surgery to the Raussendorf lattice.

To show this equivalence, we will now describe how to obtain the Raussendorf lattice from a planar code.
Figure~\ref{fig:surface_code} shows $Z$-stabilizer measurements in (a) and $X$-stabilizer measurements in (b), which were separated into two distinct time steps. Alternating between the lattices (a) and (b) and connecting data qubits in neighboring time-slices, one obtains the Raussendorf lattice. Thus, every even time-slice in the Raussendorf lattice can be associated with $Z$-stabilizer measurements and every odd time-slice can be associated with $X$-stabilizer measurements.

In the Raussendorf lattice, one can explain this procedure in terms of teleportation. The nodes in each time-slice get measured in the $X$-basis and the logical state is teleported to the next time-slice. This is the reason for the terminology of ``time-slices''.
Looking at the circuit diagrams for the planar-code stabilizer checks shown in Figure~\ref{fig:surface_code}, one can already see the similarity between the definition of graph states and these stabilizers. Each $Z$-stabilizer adds a red-colored ancillary qubit to the graph using controlled-phase gates and the parity is measured afterwards. The same happens to the $X$-stabilizers where Hadamard operations on the data qubits are required for a change in basis. These Hadamard operations are readily obtained due to the teleportation protocol described in Figure~\ref{fig:teleport1}. Thus, in each odd time-slice a Hadamard operator has been added to the data qubits and the basis transformation follows naturally from the teleportation rules.

\section{Lattice Surgery on the Raussendorf Lattice}
Now that the equivalence between the surface code and the Raussendorf lattice has been motivated, we can translate the lattice surgery protocols.
Lattice surgery is composed of many planar code patches. Each of these planar code patches encodes one logical qubit and interactions between these qubits are performed using boundary operations between neighboring patches. These boundary operations are called merge and split operations and act as parity checks.
In addition to these operations, state injection is needed to perform gates that are not supported by surface codes. With these, universality is achieved.
A description within the context of planar codes was given in the original paper for lattice surgery~\cite{Horsman2012}. Or description is different to this paper due to an additional physical dimension. This causes the planar code patches to become boxes and their edges to become faces.

The key idea of our approach is to carve out boxes from the fully-connected Raussendorf lattice which is given by the hardware. These boxes each represent a logical qubit, as did patches in the planar code. Merges occur when we stop carving out boxes and splits occur when we introduce more cuts into the lattice to split a box into separate logical boxes. Below, we describe in detail all necessary operations for universality.

\subsection{Boxes inside the Raussendorf Lattice}
  \begin{figure}
    \centering
    \includegraphics[width=\columnwidth]{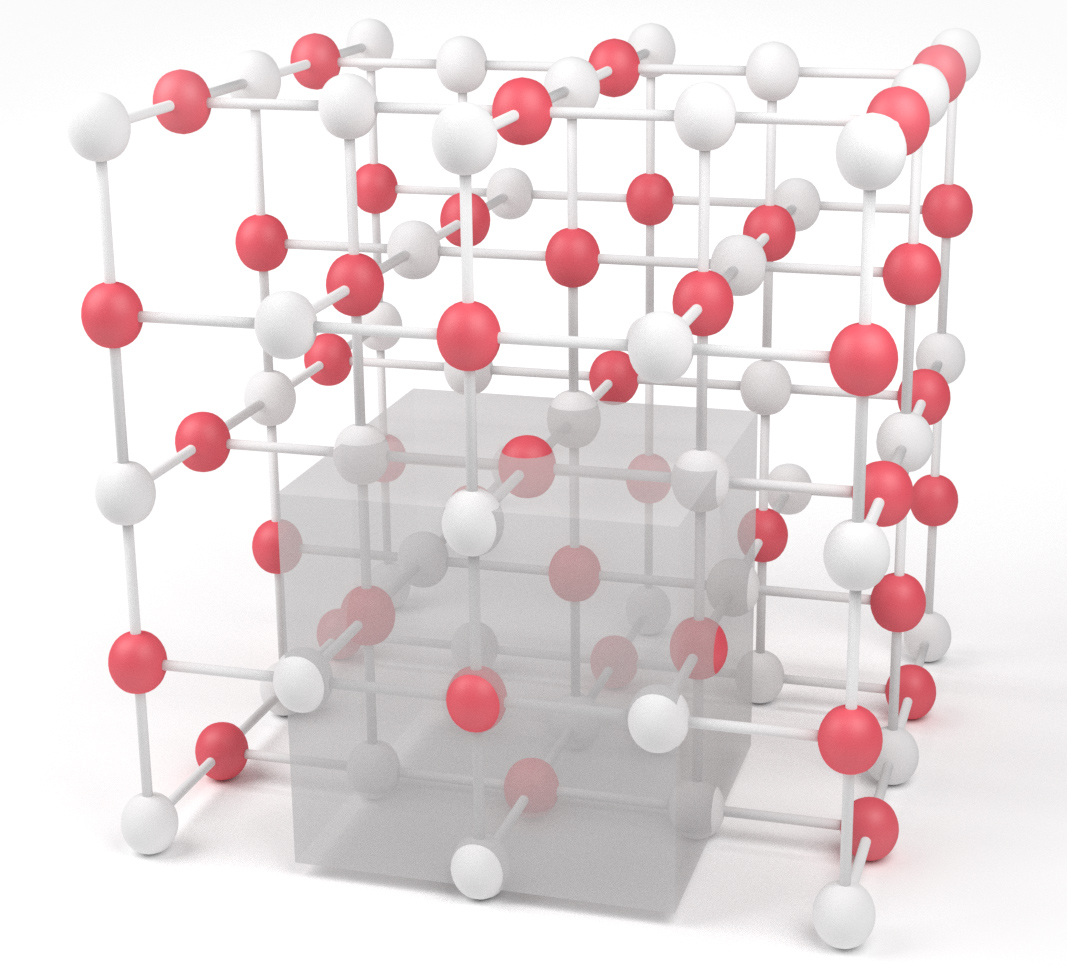}
    \caption{\label{fig:3Dpatch} 3D Representation of a Raussendorf lattice box of the patch illustrated in Figure~\ref{fig:surface_code}. The gray area corresponds to a single unit cell which was illustrated in~\ref{fig:Raussendorf_unitcell}. To separate this box from the surroundings, all qubits around the box need to be measured in the $Z$-basis.}
  \end{figure}
The qubits of the Raussendorf lattice have to be measured in the $Z$-basis if they do not belong to a logical qubit box. Conversely, nodes that contribute to a logical qubit need to be measured in the $X$-basis.
Thus, a single box inside the Raussendorf lattice can be obtained by measuring all qubits surrounding it in the $Z$-basis. In Figure~\ref{fig:3Dpatch} one such box is shown. All visible qubits are measured in the $X$-basis and the surrounding qubits have to be measured in the $Z$-basis.
There is a transition that changes $X$-measurements of the box into $Z$-measurements, which disantnagle the box from its surroundings. Depending on where this transition is, one can distinguish two types of faces: rough and smooth faces.
A rough face is cutting through primary unit cells, whereas a smooth face is composed of faces of the unit cells. From Figure~\ref{fig:3Dpatch} one can see that a single box consists of two smooth faces vertically, and two rough faces horizontally. The faces in the time direction need to be determined during initialization and measurement.

\begin{figure}
    \centering
    \includegraphics[width=\columnwidth]{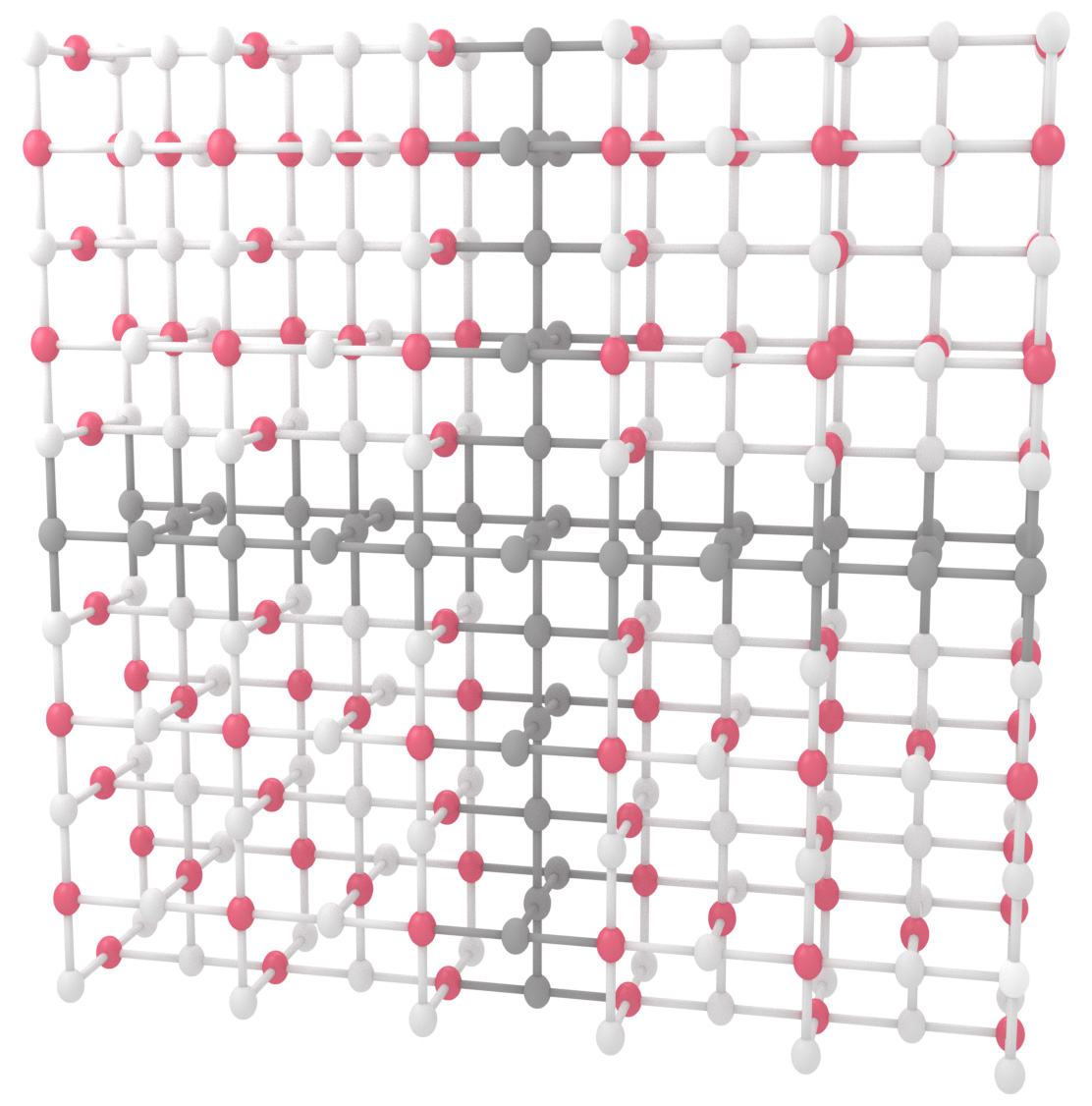}
    \caption{\label{fig:patches_array} Four planar code patches are embedded in the Raussendorf lattice with $Z$-basis measurements on the boundary between the patches (dark color).}
  \end{figure}

In Figure~\ref{fig:patches_array} we show how 4 of these patches can be arranged in the large Raussendorf lattice. The qubits on the boundary are measured in $Z$-basis, which completely disentangles each patch from the others. Boundary operations such as merges will be described later and use these qubits to add interactions between neighboring qubits.

\subsection{Logical Pauli-operators}
Logical $X$- and $Z$-operators in the original lattice surgery description can be realized by performing a physical $X$ or $Z$-operation along chains spanning from opposite edges of the patch/box. While these operations could in principle also be implemented physically, they do not need to be applied because classical software can permute these operations and reinterpret measurement results.

\begin{figure}
  \centering
  \includegraphics[width=\columnwidth]{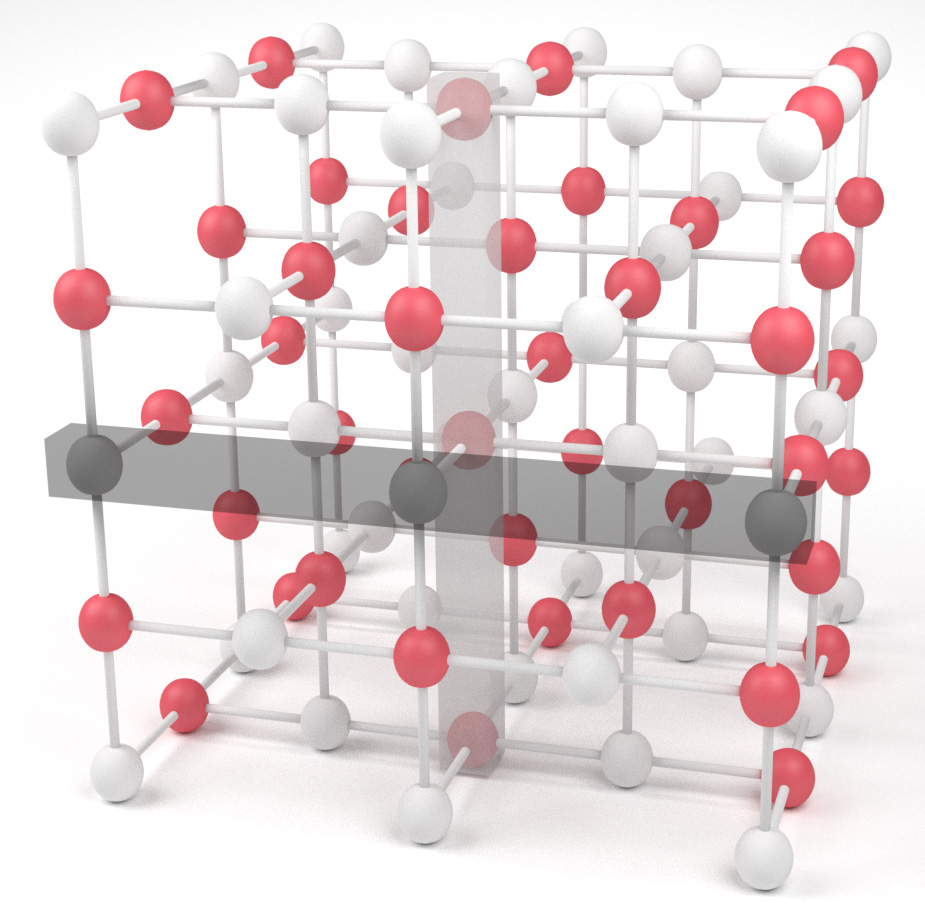}
  \caption{\label{fig:logical_operators} Logical $X$ and $Z$-operators are chains of physical $X$ or $Z$-operations. For a logical $Z$-operation this chain runs from one rough edge to the opposing rough edge (horizontally), whereas for $X$ operations it runs from smooth edge to smooth edge (vertically).}
\end{figure}

Logical $X$- or $Z$-operations can be treated the same way teleportation errors are handled on a physical qubit level. Due to by-product Pauli operations, Figure~\ref{fig:teleport1} shows, that with a probability of $50\%$, teleport operations of physical qubits have an additional error-operator. The outcome of our measurement indicates whether this error-operator was applied or not. Thus, we can treat a logical $X$-operation by inverting the measurement results along a chain on primary faces. The $Z$-operation can be performed by inverting the measurement results along a chain on a time-slice of dual faces. The logical $X$- and $Z$-operators are shown in Figure~\ref{fig:logical_operators}.

\subsection{Initialization and Measurement}
In the original description of lattice surgery~\cite{Horsman2012} patches could be initialized into the logical $\ket{0}$ state by preparing all physical qubits in $\ket{0}$. Similarly, the logical $\ket{+}$ state can be obtained by preparing all physical qubits in $\ket{+}$.
In the Raussendorf lattice a patch can be created by changing the measurements from $Z$- to $X$-basis. Therefore, these measurements stop removing qubits from the graph and start connecting subsequent qubits to each other. Due to the definition of graph states, physical qubits are always prepared in a $\ket{+}$ state. Despite this constraint, we are still able to initialize the system into the logical $\ket{0}$ or $\ket{+}$ state. The important point here is in which time-slice the measurements are changed. If the patch is created on the faces of the primal lattice (even time-slices), one will end up with a logical $\ket{+}$ state, because this would correspond to a $\ket{+}$ state initialization of each qubit.
If one chooses to switch the measurement basis on the faces of the dual lattice, each qubit will be teleported to the next primal nodes with an additional Hadamard operator (see \ref{fig:teleport1}) and therefore this will result in an initialization of the logical $\ket{0}$ state.

For measurements on logical qubits, the inverse of the initialization has to be performed. Thus, one will switch the measurement basis from $X$-measurements to $Z$-measurements. Again, the time-slice this happens determines whether an additional Hadamard gate was applied and whether the basis of the measurement was $X$ or $Z$.
For a measurement in the $Z$-basis one needs to measure an even time-slice, which corresponds to a measurement along primal faces. A measurement in the $X$-basis has to occur on an odd time-slice, which corresponds to a measurement-plane along dual faces. The measurement result can be inferred from the total parity along a chain of measurements spanning from one edge to the next. Because of the logical operators shown in Figure~\ref{fig:logical_operators} these chains need to be oriented horizontally (vertically) for a logical $Z$-measurement (logical $X$-measurement).

\subsection{Merges and Splits}
In the following we describe how merge and split operations from the original proposal~\cite{Horsman2012} can be adapted to the Rausendorf lattice.

\begin{figure}
  (a)
  \includegraphics[width=\columnwidth]{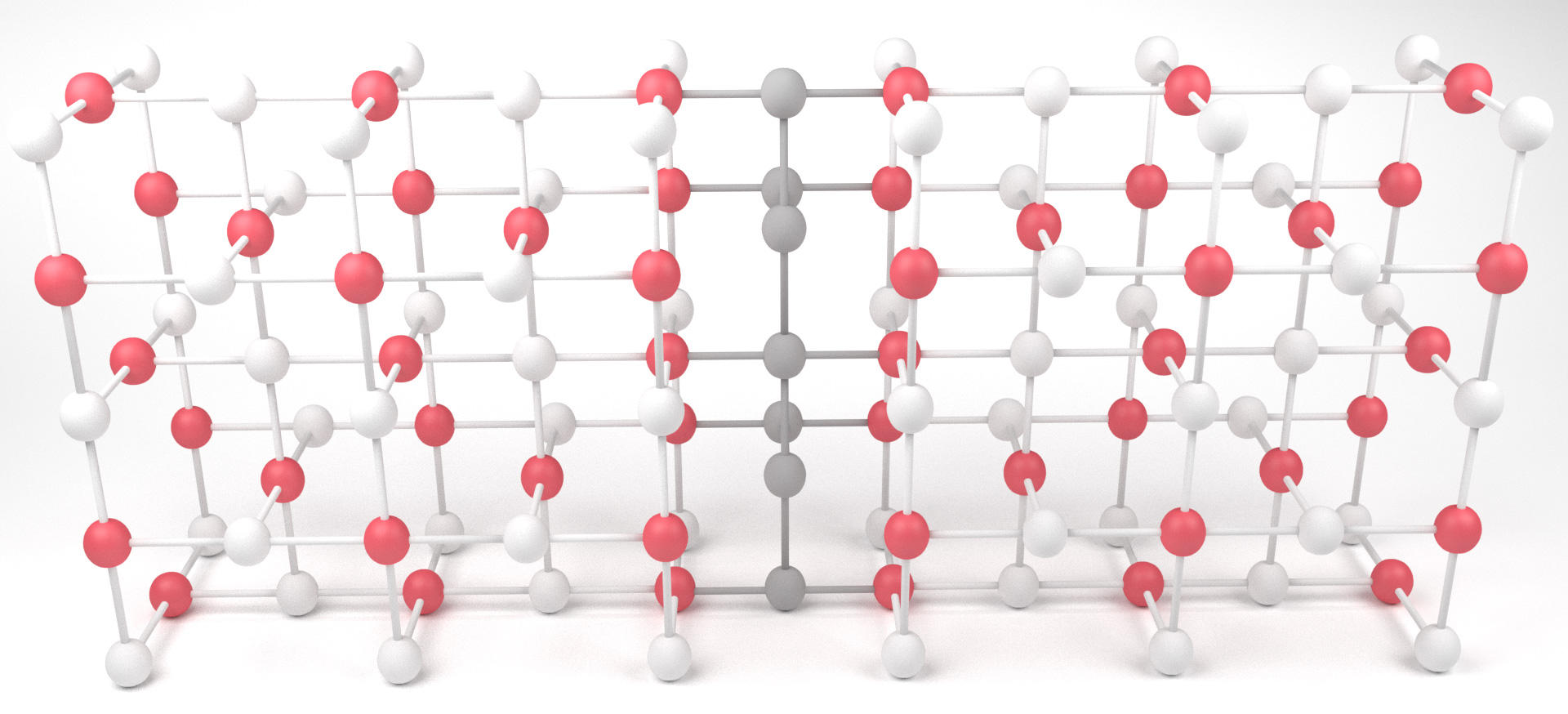}\\[0.2cm]
  (b)\\[0.1cm]
  \includegraphics[width=0.55\columnwidth]{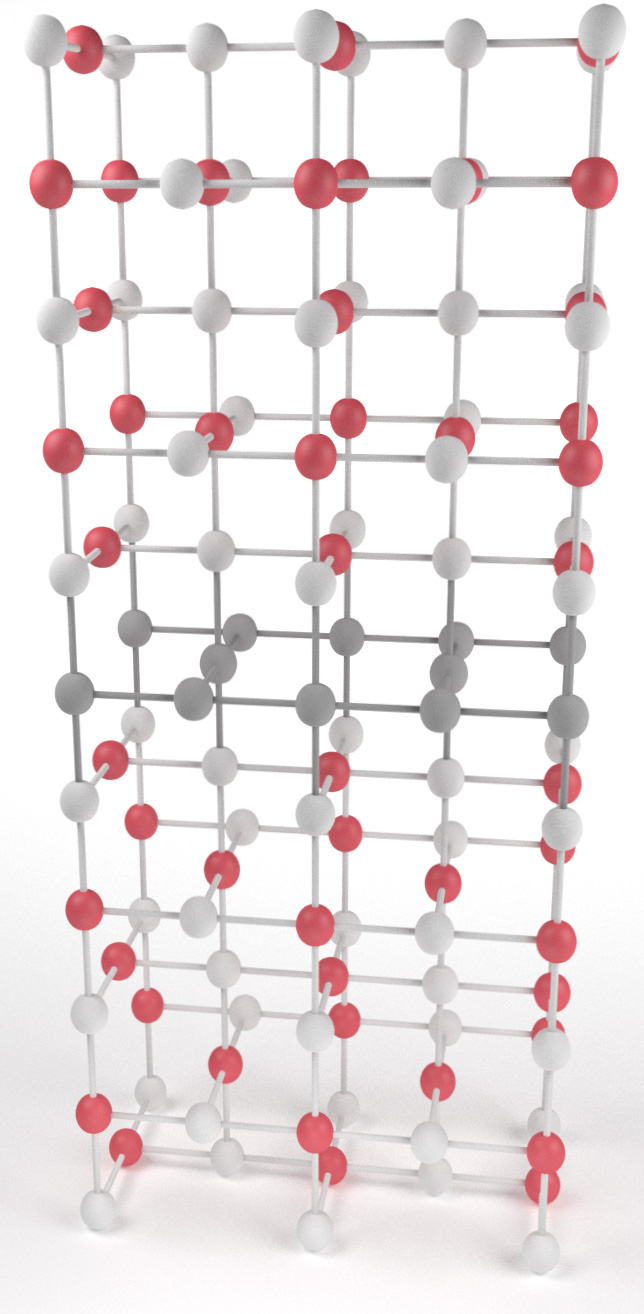}
  \caption{\label{fig:merge_operation} During a merge operation the measurement basis along the edge between two LS patches is turned on. In (a) this is shown for a rough merge, where each qubit along the boundary needs to be initialized in the $\ket{0}$ state. Thus, the measurement has to be flipped on an odd time-slice. In (b) this procedure is done for a smooth merge. Here the qubits on the boundary need to be initialized in the $\ket{+}$ state and thus the measurement basis needs to be flipped in an even time-slice.}
\end{figure}

Measurements on boundary qubits between the two boxes need to be switched from $Z$ to $X$-measurements to perform a merge operation. Data qubits on this connecting edge need to be initialized in $\ket{0}$ ($\ket{+}$) for a merge along rough (smooth) boundaries.
This requires again to swap the measurement basis at different times for smooth and rough merges. For a smooth merge, one needs to swap the basis starting at a time-slice that is made of primal faces, while for a rough merge this needs to happen on a time-slice composed of dual faces. A visualization of this can be seen in Figure~\ref{fig:merge_operation}.

The merge operation acts as a parity measurement on the two logical qubits. For a rough merge this parity corresponds to simultaneous $X$ measurement ($X X$) and for a smooth merge it corresponds to a $Z Z$ measurement on the logical states. To obtain the measurement outcome of this parity check, the faces along the boundary between the two patches have to be checked.
Because of this parity measurement, the output state is non-deterministic, and the number of logical qubits is decreased by one.

The output state of a rough merge between logical qubits $\ket{\psi} = \alpha \ket{0} + \beta \ket{1}$ and $\ket{\phi} = \alpha' \ket{0} + \beta' \ket{1}$ can be summarized in the following equation~\cite{Horsman2012}:
  \begin{align*}
    \ket{\psi} \encircle{M}_r \ket{\phi} &= \alpha \ket{\phi} + {(-1)}^M \beta \ket{\overline{\phi}} \\
    & = \alpha' \ket{\psi} + {(-1)}^M \beta' \ket{\overline{\psi}}
  \end{align*}
where $M = \left\{ 0,1 \right\}$ is the measurement outcome of the parity check and $\ket{\overline{\phi}} = \sigma_x \ket{\phi}$.

A smooth merge between between logical qubits $\ket{\psi} = a \ket{+} + b \ket{-}$ and $\ket{\phi} = a' \ket{+} + b' \ket{-}$ can be summarized in the following equation~\cite{Horsman2012}:
  \begin{align*}
    \ket{\psi} \encircle{M}_r \ket{\phi} &= a \ket{\phi} + {(-1)}^M b \ket{\overline{\phi}} \\
    & = a' \ket{\psi} + {(-1)}^M b' \ket{\overline{\psi}}
  \end{align*}
Where $M$ again gives the measurement result and $\ket{\overline{\phi}} = \sigma_z \ket{\phi}$.

\subsection{Encoding States}
  \begin{figure}
    \centering
    \includegraphics[width=\columnwidth]{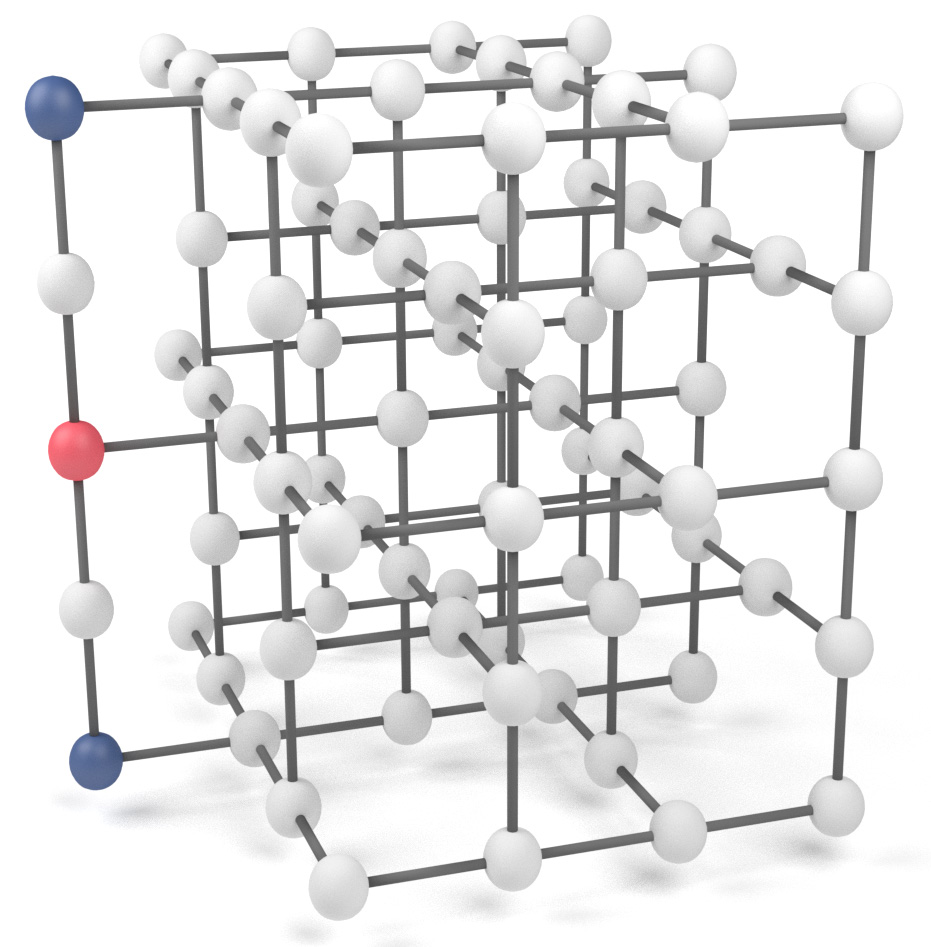}
    \caption{\label{fig:state_inject} Here, a magic state is encoded into a distance-three planar code, within the Raussendorf lattice. The time-slices are measured from the left to the right. The first measurements happen on the column that has the red-colored qubit in the middle. All qubits are measured in the standard $X$-basis, only the qubit colored in red is measured in a rotated basis. This results in a superposition state of the red and blue qubits, which is due to the measurements teleported back one layer. All other qubits need to be initialized to the $\ket{0}$ state. This is achieved by switching from a $Z$-basis to a $X$-basis measurement on an odd time-slice (dual faces).}
  \end{figure}

To be able to perform universal computation, state injection has to be performed. This requires specially prepared states as a resource. Encoding these logical states is done by using faulty single-qubit operations which require a distillation of the resulting logical state to reduce errors. Only then can these states be used as a resource for state injection. While distillation and injection itself are done on a purely logical level, a method for encoding arbitrary states needs to be devised for the Raussendorf lattice.
For injection, it is common to use the following two states, because they complement the surface code and enable universality:
\begin{align}
\begin{split}
\ket{Y} &= \frac{1}{\sqrt{2}} \left(\ket{0} + i \ket{1}\right) \\
\ket{A} &= \frac{1}{\sqrt{2}} \left(\ket{0} + e^{i\frac{\pi}{4}}\ket{1}\right)
\end{split}
\label{eq:magic_state}
\end{align}
The $\ket{Y}$ state will mainly be used for $P$-gates and Hadamard operations while the $\ket{A}$ state is needed for the implementation of the $T$-gate.
The procedure to encode logical states is based on the original lattice surgery description which we modified to be implementable in the Raussendorf lattice. It can be summarized in the following steps:
\begin{enumerate}
\item A single physical qubit is prepared as a magic state: $\ket{\psi} = \alpha \ket{0} + \beta \ket{1}$, where $\ket{\psi}$ is equal to either $\ket{A}$ or $\ket{Y}$ described in equation~\ref{eq:magic_state}.
\item Using CNOTs, a 3-qubit entangled state is created resulting in $\ket{\psi} = \alpha \ket{000} + \beta \ket{111}$
\item These three qubits are used as data qubits along a logical $Z$ operator chain for a distance-3 planar patch.
\item Then, with all other qubits of this patch initialized to $\ket{0}$, syndrome measurements are turned on.
\end{enumerate}
The resulting state of this procedure is a logical distance-three planar patch that encodes an erroneous magic state. Using merges, this patch can be brought to the desired code distance.

To implement this procedure in the Raussendorf lattice, a single qubit needs to be prepared in a magic state.
This is where a measurement in an arbitrary basis comes into play (see Figure~\ref{fig:teleport1}). Figure~\ref{fig:state_inject} shows the method of encoding an arbitrary magic state in a distance-three box. The measurement basis has to be changed from $Z$- to $X$-basis on primal faces for five qubits. Here, a vertical line of qubits is measured in the $X$-basis, except for the center qubit (indicated by the red color in Figure~\ref{fig:state_inject}). This qubit is measured in a rotated basis. In order to inject a $\ket{Y}$ state (needed for the S-gate) a rotated-basis measurement with $\theta = \pi/2$ needs to be chosen. For a T-gate, the $\ket{A}$ state has to be prepared by a rotated measurement of $\theta = \pi/4$.

The center qubit and the outer qubits (all colored qubits in Figure~\ref{fig:state_inject}) therefore teleport the state from step 2: $\ket{000} + \exp{\left(i\theta/2\pi\right)} \ket{111}$ to the next time-slice. In the second time-slice the measurement is performed in the $X$-basis for all qubits. This would correspond to an initialization of 0 for the remaining data qubits. Error correction on this distance-3 code is now possible, and using merges this state can be brought to arbitrary sizes.

This procedure encoded an arbitrary but faulty state in a logical qubit. Magic-state distillation algorithms~\cite{BravyiHaah,PhysRevX.2.041021} can now be used to reduce the noise and teleportation protocols can then implement S- and T-gates. Even, the synthillation protocols given in~\cite{synthillation} can be used here.

\subsection{Hadamard}

\begin{figure}
    \centering
    \includegraphics[width=\columnwidth]{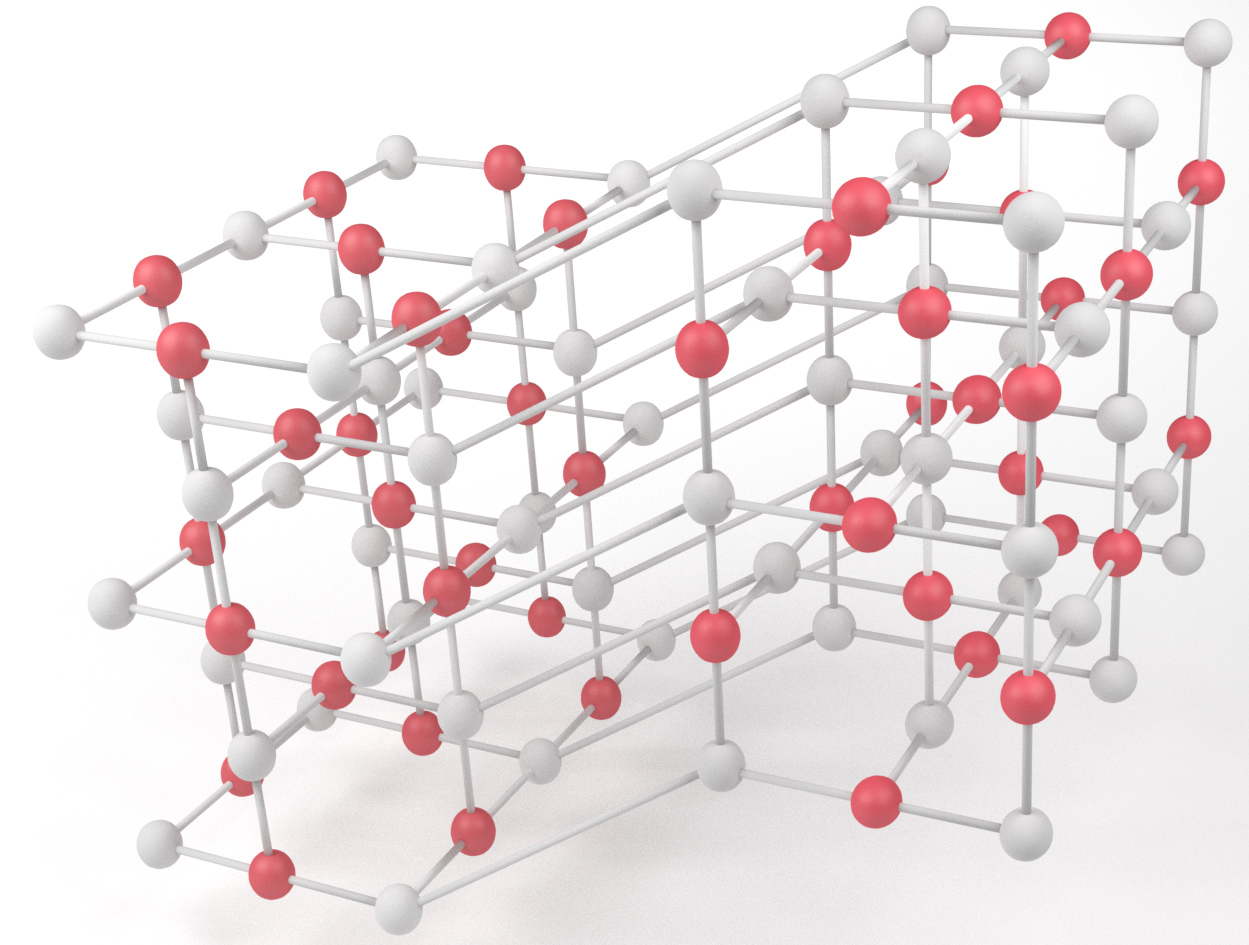}
    \caption{\label{fig:primal_primal} This Figure shows how data qubits from two primal faces can be connected. The nodes from the dual lattice are removed during the lattice generation or renormalization step. This structure archives two goals. First, it applies a Hadamard operation to each of the data qubits. Second, the data qubits are shifted by $\left(-1/2,-1/2\right)$, which allows the recombination with the rest of the Raussendorf lattice, when the Hadamard operation is finished.}
\end{figure}

A logical Hadamard operation is also possible. We present a modified version of the initial transversal Hadamard operation devised for the surface code~\cite{Fowler2012} and lattice surgery~\cite{Horsman2012}. However, in order to recombine the modified logical state with its surroundings, we need to employ operations that are not native to the Raussendorf lattice itself. Looking at the lattice generation protocol of~\cite{renormalization} we can modify the lattice by choosing a different method for the lattice renormalization.

Our proposal for a logical Hadamard operation starts with a physical Hadamard operation applied to each data qubit.
This is achieved by connecting primal nodes with primal nodes as shown in Figure~\ref{fig:primal_primal}. Due to the missing temporal layer of nodes, an additional by-product Hadamard is applied. Furthermore, data qubits are shifted by half a unit cell down and left. This is required in order to be able to recombine the lattice with other patches later on.

In comparison to Figure~\ref{fig:logical_operators}, their logical operators have now changed the measurement basis for physical qubits. This means that the logical $Z$-operator is a horizontal chain of physical $X$-operations instead of physical $Z$-operations. The logical $X$-operator is now described by a chain of vertical $Z$-operations.

\begin{figure}
    \centering
    \includegraphics[width=\columnwidth]{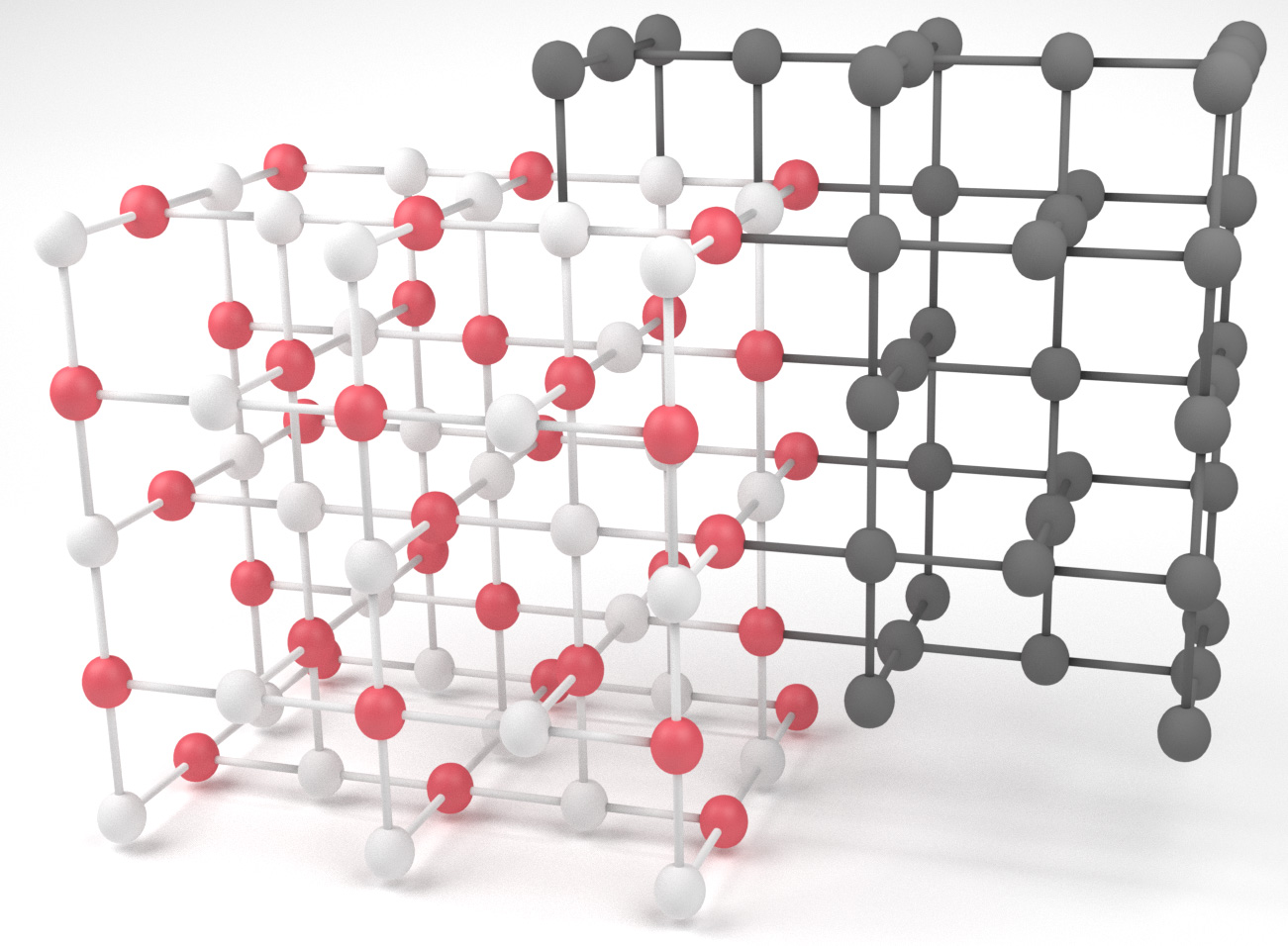}
    \caption{\label{fig:rotation} This procedure rotates the patch by $\pi/2$. The chains of logical operators are rotated such that they are facing the same direction as logical operators of other patches.}
\end{figure}

To complete the Hadamard translation, this patch has to be rotated. In the original lattice surgery proposal~\cite{Horsman2012} such a rotation has already been proposed and can simply be translated to the Raussendorf lattice. In Figure~\ref{fig:rotation} this rotation is performed by switching the measurement basis on dual faces from $Z$ to $X$ for all qubits depicted in dark-grey (This corresponds to a $\ket{0}$-initialization for qubits on the surface code).

The resulting patch is still a distance-$d$ planar code. After $d$-rounds of error correction all qubit measurements that do not form a square with the dark-colored ones can be turned to $Z$-basis measurements. This results in a rotated patch, such that a previous horizontal logical operator is now vertical and vice versa. Additionally, this rotation cancels the shift by half a unit cell that was introduced before, such that the patch is back in proper alignment with the rest of the Raussendorf lattice.

The logical $Z$-operator, that previously was a horizontal chain of physical $Z$-operators, has now been mapped to a chain of vertical $X$-operations. Thus, the logical $Z$ and $X$ operators have been swapped and a proper Hadamard operation has been applied.

\subsection{Hadamard and ICM}
\begin{figure}
    \centering
    \includegraphics[width=0.49\columnwidth]{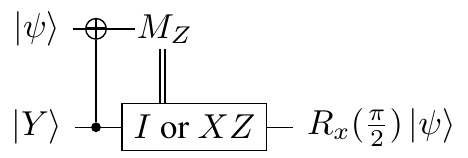}
    \includegraphics[width=0.49\columnwidth]{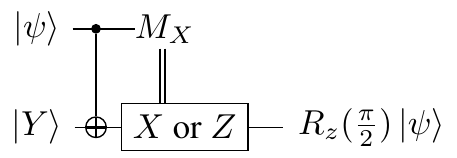}
    \caption{\label{fig:teleport2} Teleportation operations needed to implement a Hadamard operation. The circuit to the left implements a $X$-rotation and the circuit to the right implements a $Z$-rotation.}
\end{figure}

For some hardware models, connections between primal nodes might not be allowed. Thus, we present another approach which is based on the decomposition of a Hadamard into rotational gates $H = R_z\left(\frac{\pi}{2}\right)R_x\left(\frac{\pi}{2}\right)R_z\left(\frac{\pi}{2}\right)$. These rotational gates can be implemented using the measurement procedure shown in Figure~\ref{fig:teleport2}, which treats the application of Hadamards on a purely logical level.

Unfortunately, this procedure needs state injection and is more costly due to the distillation protocols that are required. Once the necessary $\ket{Y}$ states have been encoded and distilled, the teleportation protocols given in Figure~\ref{fig:teleport2} will implement the rotation operations required by the Hadamard operation.

A convenient representation that translates any circuit using teleportation gates into a deterministic circuit is the ICM model~\cite{Braiding_compiler,ICM_orig}, which stands for (I)nitialization, (C)NOT and (M)easurements. In this representation, any operation is decomposed into teleportation operations, and non-deterministic results are handled using selective target and selective source teleportation methods~\cite{PDF16,ICM_orig}. This results in a deterministic circuit, despite being dependent on the probabilistic outcome of its teleportation operations.

A further extension of this representation is the inverted ICM representation~\cite{herr_lattice_surgery}. The difference between ICM and inverted ICM is that the former has arbitrary basis initialization and restricted-basis measurements, whereas the latter has only a restricted initialization ($Z$ or $X$) and arbitrary basis measurements. In the inverted ICM representation, an error-corrected graph state which is specific to the algorithm is created and can be readily realized by merges and splits of lattice surgery. Afterwards, measurements perform the computation of the quantum algorithm. A complete discussion is available in~\cite{herr_lattice_surgery}.

\section{Conclusion}
So far the literature has only described braiding as a method of computation on the Raussendorf lattice. In this paper we showed that a lattice surgery implementation can also perform quantum computation in the Raussendorf lattice and we described how to implement all fundamental operations.
An implementation of transversal Hadamards proved to be hard without changing the lattice structure. We gave two approaches, both with different drawbacks. In conclusion, lattice surgery can be implemented on the Raussendorf lattice such that future quantum computation~\cite{rudolph} on the Raussendorf lattice can use the beneficial aspects of lattice surgery.

\begin{acknowledgments}
D.H. is supported by the RIKEN IPA program and A.P. is supported by the Linz Institute of Technology project CHARON, grant number LITD13361001.

S.J.D. acknowledges support from the JSPS Grant-in-aid for Challenging Exploratory Research and from the Australian Research Council Centre of Excellence in Engineered Quantum Systems EQUS (Project CE110001013).

FN is partially supported by the MURI Center for Dynamic Magneto-Optics via the AFOSR Award No.~FA9550-14-1-0040,
the Army Research Office (ARO) under grant number 73315PH,
the AOARD grant No.~ FA2386-18-1-4045,
the CREST Grant No.~JPMJCR1676,
the IMPACT program of JST,
the RIKEN-AIST Challenge Research Fund,
the JSPS-RFBR grant No.~17-52-50023,
and the Sir John Templeton Foundation.

\end{acknowledgments}
\bibliography{biblio}

\end{document}